# Fast heat flux modulation at the nanoscale

P.J. van Zwol[1*], K. Joulain[2], P. Ben Abdallah[3], J.J. Greffet[3], J. Chevrier[1]

[1]*Institut Néel, CNRS and Universite Joseph Fourier Grenoble, BP 166 38042 Grenoble Cedex 9, France*
[2]*Institut P', CNRS-Université de Poitiers-CNRS UPR 3346, 86022 Poitiers Cedex, France*
[3]*Laboratoire Charles Fabry, Institut d'Optique, CNRS, Université Paris-Sud, Campus Polytechnique, RD 128, 91127 Palaiseau cedex, France*

*Electronic address    petervanzwol@gmail.com

We introduce a new concept for electrically controlled heat flux modulation. A flux contrast larger than 10 dB is expected with switching time on the order of tens of nanoseconds. Heat flux modulation is based on the interplay between radiative heat transfer at the nanoscale and phase change materials. Such large contrasts are not obtainable in solids, or in far field. As such this opens up new horizons for temperature modulation and actuation at the nanoscale.

Pacs numbers; 44.40.+a,  44.05.+e, 64.70.Nd

While ultrafast modulators of light fluxes or electrical current are standard devices routinely used for information processing, heat flux modulators are not easily available. There is a fundamental reason for this striking difference: perfectly opaque materials or perfect electrical insulators exist but perfect thermal insulators do not exist. Whereas optical transmission and electrical resistances can be modulated over several decades, heat conductivity varies between 200 $Wm^{-1}K^{-1}$ for good conductors such as aluminum, and 0.15 $Wm^{-1}K^{-1}$ for a poor conductor such as asbestos and 0.02 $Wm^{-1}K^{-1}$ for dilute materials such as air. Yet, controlling heat fluxes could be of great interest for many purposes. Many materials properties depend on the temperature, chemical and biochemical reactions are very sensitive to temperature, thermal energy management is becoming a key issue in nanoelectronics. Hence, it would be extremely interesting to have a device permitting to open and close a heat transfer channel at a high speed rate. Some steps have been taken along this direction. The concept of thermal transistor has been introduced recently in the context of low temperature physics and it was shown that the heat flux of a mesoscopic system operating in the 100-500 mK range can be electrically controlled [1]. Numerical simulations of one dimensional chains of non linear oscillators have been shown to mimic a transistor [2]. This work is related to the non-linearity at work in thermal rectifiers [2, 3]. Yet, no electrically controlled heat transfer modulators operating at ambient temperature exist so far. Here, we introduce a new way of modulating radiative heat flux. Our method works in high vacuum at room temperature, and requires only solid state materials. We show that modulations larger than 10 dB can be obtained with a switching time as small as 10 ns.

The working principle is based on the radiative heat transfer at the nanoscale [4-13]. It has been demonstrated recently that the flux between two plane parallel planes can be enhanced by orders of magnitude due to the contribution of evanescent waves [4-7]. More specifically, it has been predicted [7] and verified experimentally [5, 6] that the contribution of surface phonon polaritons increase the flux by more than an order of magnitude. The mechanism at work in this heat transfer flux is the dipole-dipole interaction [13] so that only the atoms located very close to the surface contribute to the flux. It follows that by modifying the properties of a thin layer, the flux can be dramatically modified.

In order to modulate the heat flux, we build on previous experience in the field of Casimir forces. Both thermal transfer and Casimir force depend on the dielectric properties of the material [4-21]. By using Phase Change Materials (PCM), modulations of the Casimir force on the order of 25 % have been demonstrated [15]. There exists a large dielectric contrast between the two phases, which has been recently attributed to a change of bonding on crystallization [22]. Phase Change Materials are interesting as they can be switched between an amorphous and a crystalline state with a switching time on the order of a few nanoseconds [22-25]. We emphasize that the modulation of the flux is due to an electrically controlled phase change so that it can be very fast. It is not limited by the thermal inertia of the system.

In what follows, we introduce the basics of heat transfer at the nanoscale and the optical and IR properties of PCM. We then explore the value of the heat flux for different systems including thin layers of PCM. We propose a gold/glass system with a thin layer of PCM as a candidate for modulating the heat transfer. Finally, we discuss a potential application of the modulator for mechanical actuation.

The calculations are done for planar parallel surfaces. For a detailed account of the theory, the reader is referred to refs. [6, 17-19]. The theory is based on the fluctuation-dissipation theorem that is applied to Maxwell equations augmented with stochastic sources [12]. Here it suffices to state that we numerically integrate equation 23 of reference [19] to obtain the heat transfer $q_{net}$ between two surfaces at temperatures $T_1$ and $T_2$

$$q''_{net} = \int_0^\infty \frac{d\omega}{2\pi}[\Theta(w,T_1) - \Theta(w,T_2)] \times$$
$$\left\{ \int_{q<\omega/c} \frac{d^2q}{(2\pi)^2}\left[\frac{[1-|R_{1p}(q,\omega)|^2]\cdot[1-|R_{2p}(q,\omega)|^2]}{|1-e^{i2\gamma d}R_{1p}(q,\omega)R_{2p}(q,\omega)|^2}\right] + \right.$$
$$\int_{q>\omega/c} \frac{d^2q}{(2\pi)^2}e^{-2|\gamma|d}\left[\frac{\text{Im}\,R_{1p}(q,\omega)\,\text{Im}\,R_{2p}(q,\omega)}{|1-e^{-2|\gamma|d}R_{1p}(q,\omega)R_{2p}(q,\omega)|^2}\right]$$
$$\left. + [p \to s] \right\} \quad , (1)$$

where $\Theta(\omega, T) = \hbar\omega/[\exp(\hbar\omega/k_bT)-1]$ is the mean energy of a Planck oscillator at frequency $\omega$ in thermal equilibrium and $\gamma = \sqrt{[(\omega/c)^2 - q^2]}$. The symbol $[p \to s]$ denotes the first two terms for which the p-polarized reflection coefficients $R_p(q,\omega)$ are replaced by the s-polarized coefficients $R_s(q,\omega)$. The difference $\Theta(\omega, T_1) - \Theta(\omega, T_2)$ can be linearized for $T_1 - T_2 \ll T_1$ so that the heat flux can be cast in the form $q = h(T_1 - T_2)$ where h is the heat transfer coefficient. Equation 1 is valid in both near field ($q > \omega/c$) and far field ($q < \omega/c$) as long as non-local effects are neglected i.e. when distances are larger than the atomic dimensions for dielectrics (1nm) or the Thomas-Fermi screening length for metals [20,21]. Equation (1) depends on the dielectric properties of the surfaces. Dielectric data are obtained for gold [14,27], glasses [16, 26] and PCM surfaces [15], from the far infrared (IR) to the ultraviolet spectral ranges. Further details can be found in the references.

In what follows, we show how the heat transfer between two parallel plates depends on the gap width. We will then study the influence of the dielectric properties of the PCM material. Figure 1 displays the flux between two gold surfaces. It is seen that the flux is enhanced by three orders of magnitude when the gap width is reduced from some µm to 100 nm. We have computed the flux for different values of the dielectric data sets to illustrate the influence of the uncertainty. The calculations are quite sensitive to the fitted Drude parameters. For example in the case of Palik's handbook [27] data, the measured range was limited to a wavelength of 10µm. In this case the Drude model fit gave a much larger relaxation frequency as compared to the other films for which data was available up to 33µm, thus affecting heat transfer calculations as well. For gold surfaces, the uncertainty on the dielectric constant reaches 50% in the worst case for the different gold samples. Consequently the variations in radiative heat transfer for the different samples is about 40% (figure 1).

For two glass surfaces, we observe an even larger enhancement of the heat flux at short distance (figure 2). It is due to the contribution of surface phonon polaritons. Again, we study the variation of the heat transfer when accounting for uncertainty on dielectric data. The variation in heat transfer is only 5% for the two silica samples. When instead, such as in reference [6] borosilicate or sodalime glass is used, which may have a slightly different absorption spectrum (as can be derived from the upper inset in fig.2) the difference in heat transfer may be 20% as compared to a silica sample. Note here that for borosilicate glass no absorption spectrum was available for frequencies below $10^{14}$rad, thus the spectrum for sodalime glass was used to cover that range.

We now turn to the heat flux when using PCM materials. They exhibit a very high dielectric or conductivity contrast between their amorphous and crystalline states: the crystalline state behaves like a metal, whereas the amorphous state behaves more like a dielectric or semiconductor. As an example we will use a material called AIST, an alloy of silver, indium, antimony and tellurium. It is already commercially used for memory cells. The material can be crystallized, or re-amorphised in a few nanoseconds using a current pulse that heats the material. Heating above the melting temperature induces a transition phase towards the amorphous state. A longer heating pulse allows the recrystallisation. After the pulse, regardless of the state, the PCM material remains in the same state for years. Thus the phase of the cell does not depend on the temperature to a large extent. PCM can currently be reversibly switched during $10^7$-$10^{12}$ cycles [24] depending on the configuration (linecell or ovonic type). The dielectric function of AIST was measured and analyzed in ref [15] for the wavelength range 130nm to 33 micron by employing ellipsometry. For the wavelength range above 33μm a Drude model was fitted to the crystalline film, whereas the amorphous film was modelled like a dielectric (negligible or very low far IR absorption). For both phases there may be some uncertainties in the thermal transfer calculations because we do not know exactly how these materials absorb in the wavelength range beyond 33 micron. In any case these uncertainties will not matter much, since the part of the flux that exists in the range covered by the ellipsometry measurements between 130 nm and 33 μm, defines 95% of the total heat flux in case of the amorphous state, and 70% in case of the crystalline state.

Radiative heat transfer calculations for these materials are shown in Figure 3. The calculations were done for PCM-gold and PCM-glass surfaces. It is seen that the heat flux strongly depends on the phase of the PCM. The highest heat transfer contrast between the two phase states is obtained for the PCM-Gold system. The ratio between the two cases can be larger than an order of magnitude. Whereas in case of the glass-PCM system, the net heat transfer is higher as expected due to the surface phonon polariton contribution [7], the contrast due to the phase change is somewhat smaller with a factor of 8 for a gap width of 10nm. For both gold and glass, the crystalline state gives the largest heat transfer at small separations. Interestingly however, in the case of PCM-Glass, the situation reverses for separations larger than 100nm, where the amorphous state gives the highest heat transfer. The crossover also exists for gold but is less visible. This happens due to the fact that the strong resonances in near field always correspond to a frequency range where reflection is strong in far field. For the amorphous state, the reflection coefficient hardly reaches values close to one whereas this is typical for the crystaline state. In the near field the resonances for the crystalline state are much stronger than for the amorphous state.

We now examine in more detail the possibility of practical heat flux modulation. For PCM to be switched by a current, it must be protected by a capping layer, usually a glassy non conductive material. In order to achieve a rapid switching, we need to use a thin layer. We report calculations for a thin layer of PCM deposited on a $SiO_2$ substrate with a 5nm $SiO_2$ capping layer. For such multilayer systems, it suffices to replace the reflection coefficients that go into in eq. (1) [19] with those of the multilayer system. For a glass probe that hovers above this system the contrast was smaller than 50%. However for a gold probe the contrast is found to be as large as 32 as shown in figure 4. With a PCM layer of 10nm thickness, the heat transfer contrast was still a factor of 5. For such thin layers the conductivity contrast is conserved, and the switching speed is improved [25].

Let us emphasize that the flux can be either positive or negative. In other words, by increasing the conductance between a sample and a colder medium, we can remove heat from the sample. Alternatively, the flux could be modulated by changing the gap width but this could be done only with moving pieces and hence, with longer switching times. Note that typically the thickness of a PCM layer may change by 5% upon phase transition. For a typical PCM cell with thickness 30nm this amounts to a change in distance of only 1.5nm. This does not severely affect the working of a real device.

Obviously, the phase change also induces a thermal conductivity change in the PCM material. It changes by a factor of four [28] which is significant, but almost an order of magnitude lower than the contrast observed here. However, as PCM is necessarily deposited on a substrate, one has to take into account other resistances in series and parallel when evaluating the overall conductance modulation. Hence, the overall heat transfer modulation is significantly reduced. In the near-field configuration, the radiation resistance is in series with the substrate. Since it is the largest resistance, its modulation result in a large modulation of the overall resistance.

In summary, we have shown that by combining radiative heat transfer at the nanoscale with phase change materials [22-25], heat transfer can be modulated fast with a switching time on the order of a few nanoseconds. PCM has a proven cyclability that scales up to $10^{12}$ cycles [24], furthermore the switching properties remain good for even the smallest cells [25]. This ensures the realization of a practical heat switch. The unique large thermal conductivity contrast observed here cannot be obtained in solids, or in the far-field. As such this opens up new ways of fast thermal control of nanoscale devices. For example it may be used as a way of controlled cooling at the nanoscale, for fast switching thermal transistors, or in thermally activated mechanical systems. Finally there may be further room to enhance the contrast by carefully choosing the materials.

**Acknowledgements**

We would like to acknowledge useful discussions with J. Drevillon and E. Nefzaoui. We gratefully acknowledge support of the Agence Nationale de la Recherche through the Source-TPV project ANR 2010 BLAN 0928 01.

**References**

1. O. P. Saira, et. al, Phys. Rev. Lett. **99**, 027203 (2007).

2. B. Li, L. Wang, and G. Casati, Appl. Phys. Lett. **88**, 143501 (2006); ibid, Phys. Rev. Lett. **93**, 184301 (2004); M. Terraneo, M. Peyrard, and G. Casati, Phys. Rev. Lett. **88**, 094302 (2002)

3. C. R. Otey, W. T. Lau, and S. H. Fan, Phys. Rev. Lett. **104,** 154301 (2010).

4. A. Kittel, W. Müller-Hirsch, J. Parisi, S. A. Biehs, D. Reddig, M. Holthaus, *Phys.Rev. Lett.* **95**, 224301 (2005).

5. S. Shen, A. Narayanaswamy, and G. Chen, Nano Lett. **9**, 2909 (2009).

6. E. Rousseau, A. Siria, G. Jourdan, S. Volz, F. Comin, J. Chevrier, and J-J. Greffet, Nat. Photonics **3**, 514 (2009).

7. J.-P. Mulet, K. Joulain, R. Carminati, and J.-J. Greffet, Appl. Phys. Lett. **78**, 2931 (2001).

8. F. Intravaia and A. Lambrecht. Phys. Rev. Lett. **94**, 110404 (2005)

9. C. Henkel, K. Joulain, J.-Ph. Mulet, and J.-J. Greffet, Phys. Rev. A **69**, 023808 (2004).

10. E. G. Cravalho, C. L. Tien, and R. P. Caren, *J. Heat Transfer* **89**, 351 (1967).

11. A. Olivei, *Rev. Phys. Appl.* **3**, 225 (1968).

12. D. Polder and M. Van Hove, *Phys. Rev. B* **4**, 3303 (1971).

13. G. Domingues, S. Volz, K. Joulain, and J.-J. Greffet, Phys. Rev. Lett. **94**, 085901 (2005)

14. V. B. Svetovoy, P. J. van Zwol, G. Palasantzas, and J. Th. M. De Hosson, *Phys. Rev. B* **77**, 035439 (2008).

15. G. Torricelli et al, Phys. Rev. A. Rapid Comm. 82 010101 (2010).

16. P. J. van Zwol, G. Palasantzas, and J. Th. M. DeHosson, Phys. Rev. E **79**, 041605 (2009).

17. K. Joulain, J.-P. Mulet, F. Marquier, R. Carminati, and J.-J. Greffet, *Surf. Sci. Rep.* **57**, 59 (2005)

18. C. J. Fu and Z. M. Zhang, Int. J. Heat Mass Transfer **49**, 1703 (2006)

19. A. I. Volokitin and B. N. J. Persson, Rev. Mod. Phys. **79**, 1291 (2007).

20. C. Henkel and K. Joulain, Appl. Phys. B: Lasers Opt. **84**, 61 (2006).

21. P. O. Chapuis, S. Volz, C. Henkel, K. Joulain, and J. J. Greffet, Phys. Rev. B **77**, 035431 (2008)

22. K. Shportko et al, *Nat. Mater.* **7**, 653 (2008).

23. R. Pandian et al, Adv. Mater. **19**, 4431 (2007).

24. M. H. R. Lankhorst, B. W. S. M. M. Ketelaars, and R. A. M. Wolters, Nat. Mater. **4**, 347 (2005).

25. G. Bruns et al, Appl. Phys. Lett. **95**, 043108 (2009).

26. R. Kitamura, L. Pilon, and M. Jonasz, Appl. Opt. **46**, 8118 (2007)

27. E. D. Palik, *Handbook of Optical Constants of Solids* (Academic, New York, 1995).

28. X jao, et al, Appl. Phys. A 94, 627–631(2009)

29. K. L. Ekinci, Small 1, 786-797 (2005)

30. M. C. LeMieux, M. E. McConney, Y.-H. Lin, S. Singamaneni, H. Jiang, T. J. Bunning, and V. V. Tsukruk, Nano Lett. **6**, 730 (2006).

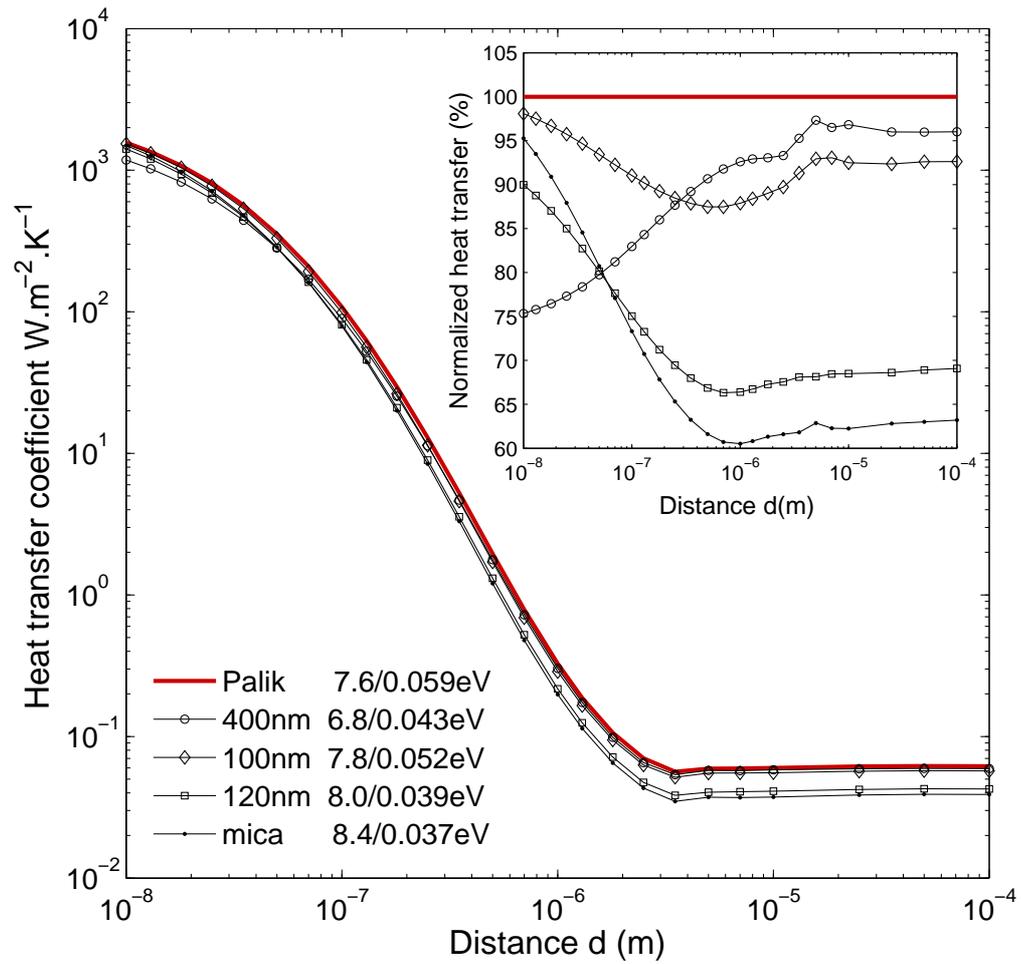

Figure 1: (color online) Radiative heat transfer for different gold surfaces. The inset shows heat transfer normalized for the Palik case, such that the differences become more visible. The Palik handbook data is compared to gold films with different thicknesses deposited on Si. Also data for a 120nm thick annealed film on mica is shown. The Drude parameters are indicated.

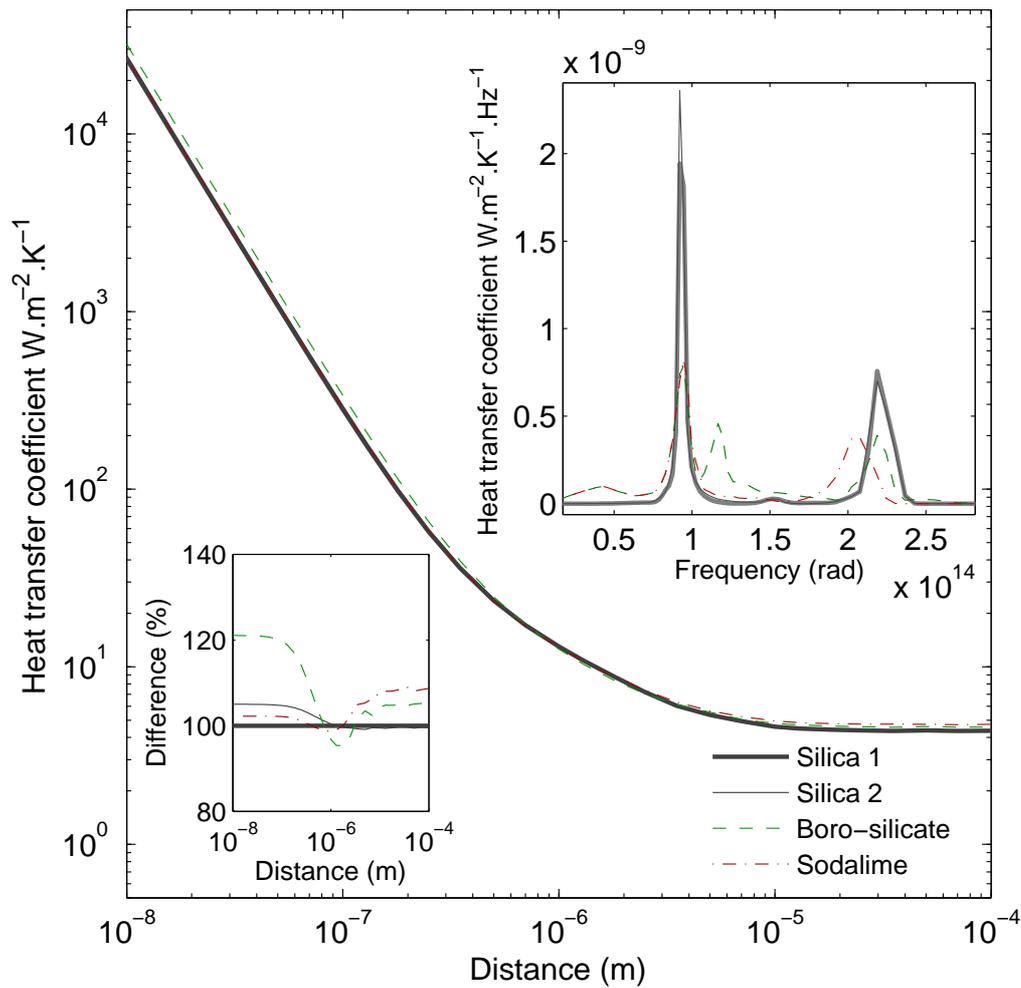

Figure 2: (Color online) Radiative heat transfer for different glass surfaces. The lower inset shows heat transfer normalized for one silica sample, such that the differences become more visible. The upper inset shows the spectral heat transfer coefficient for the different types of glass at a distance 10nm.

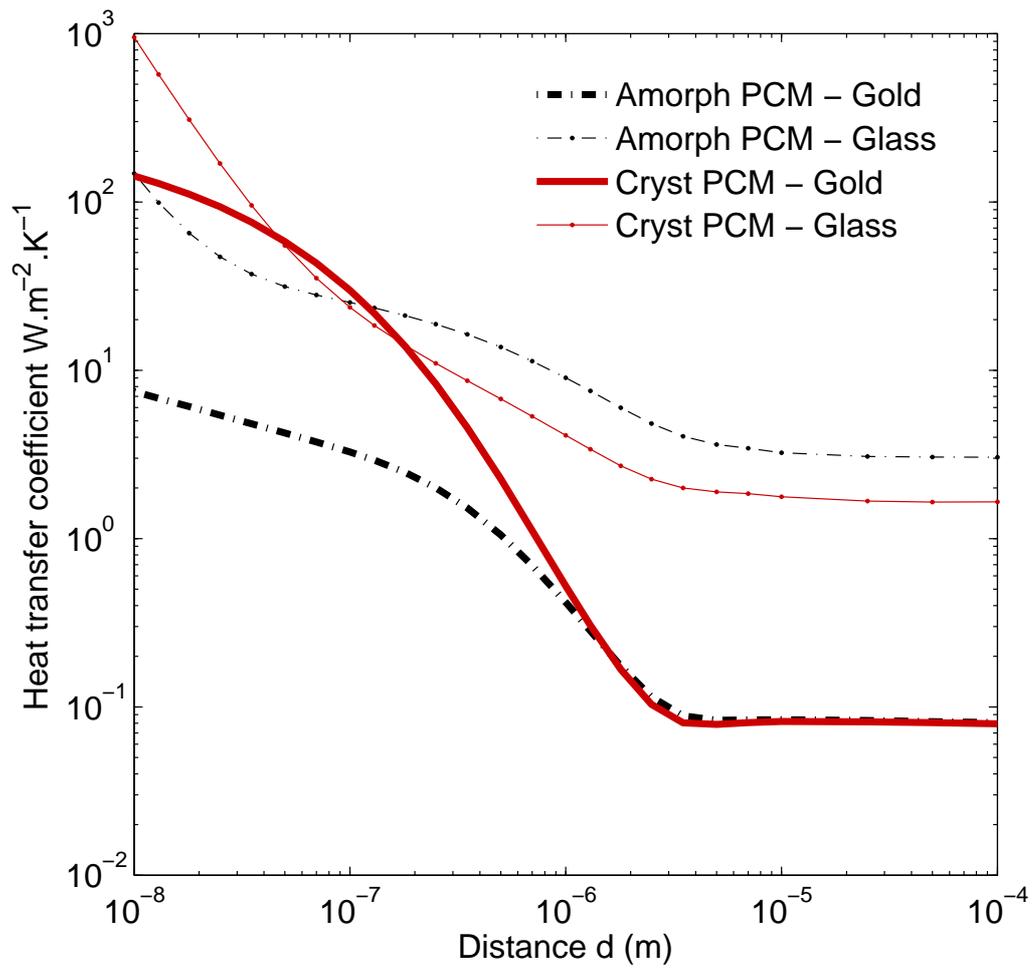

Figure 3: (Color online) Radiative heat transfer for PCM-glass and PCM-gold.

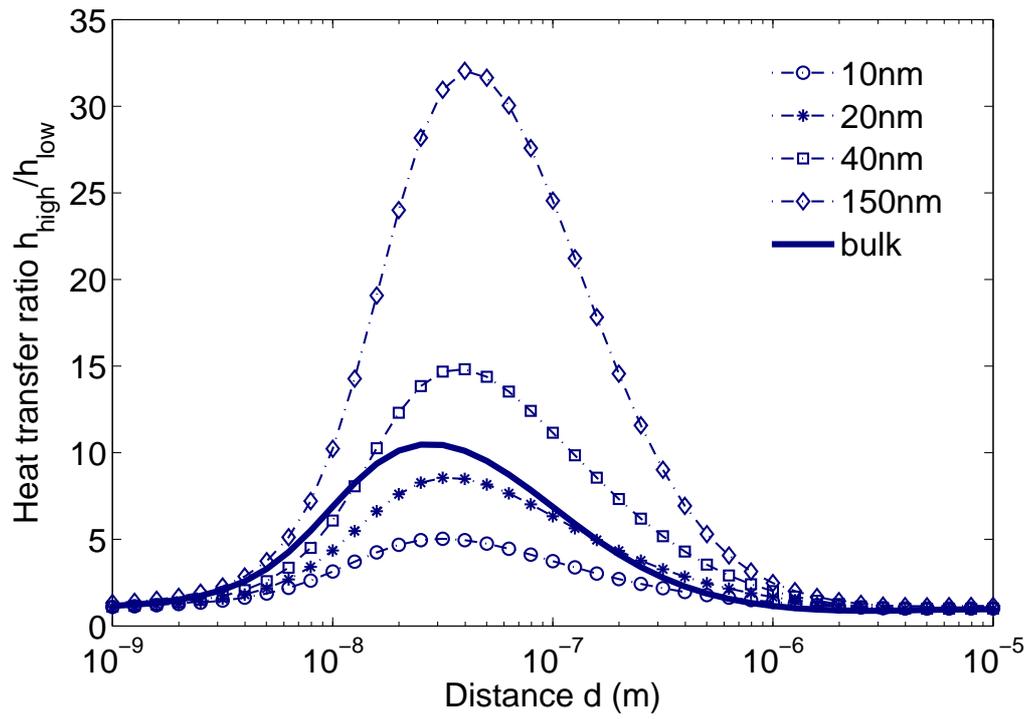

Figure 4: (Color online) Modulation of heat transfer in vacuum between a gold surface and a switchable PCM (AIST) layer on a glass substrate protected by 5nm glass top layer. Data are shown for different thicknesses of the PCM material.